\RequirePackage{fix-cm}
\documentclass[twocolumn]{svjour3}          
\smartqed  
\usepackage{graphicx}
\journalname{Microfluidics and Nanofluidics}
\begin{document}

\title{Single-pulse dynamics and flow rates of inertial micropumps}




\author{A.N. Govyadinov           \and
        P.E. Kornilovitch         \and
        D.P. Markel               \and  
        E.D. Torniainen }


\institute{HP Inc.                         \\
              Corvallis, Oregon 97330, USA \\
              Tel.: +1-541-715-0895        \\
              \email{alexander.govyadinov@hp.com}           
}

\date{Received: date / Accepted: date}

\maketitle

\begin{abstract}

Bubble-driven inertial pumps are a novel m\-e\-thod of moving liquids through microchannels. We co\-mbine high-speed imaging, computational fl\-u\-id dynamics (CFD) simulations and an effective one-dimensional mo\-del to study the fundamentals of inertial pumping. For the first time, single-pulse transient flow through U-shaped microchannels is imaged over the entire pump cycle with 4-$\mu$s temporal resolution. Observations confirm the fundamental N-shape flow profile predicted earlier by theory and simulations. Experimental flow rates are used to calibrate the CFD and one-dimensio\-nal models to extract an effective bubble strength. Then the frequency dependence of inertial pumping is studied both experimentally and numerically. The pump efficiency is found to gradually decrease once the successive pulses start to overlap in time.      



\end{abstract}

\section{\label{frdep:sec:one}
Introduction
}

Continuing development of microfluidics requires reliable integrated components such as micropumps, microvalves, micromixers and microheaters~\cite{Laser2004,Oh2006,Garimella2006,Nabavi2009,Lee2011}. Of particular interest are devices with no moving parts because of their relative robustness and ease of fabrication. Such dynamic devices have to operate under different principles than their more conventional mechanical counterparts. One approach has been to utilize high-pressure vapor bubbles as mechanical actuators. The bubbles can be created by compact microheaters embedded inside microchannels and they have enough power to displace fluid in channels as narrow as several microns across. The challenge then is to convert displacements into useful microfluidic functions. 

Several bubble-driven micropump designs have been proposed. Some of them utilize asymmetric nozzle-dif\-fu\-ser geometries~\cite{Tsai2002,Jung2007} while others use multiple heaters to create asymmetric heating~\cite{Takagi1994,Ozaki1995} or a traveling vapor plug~\cite{Jun1998,Song2001,Yokoyama2004,Okuyama2005,Bezuglyi2007}. A more recent development is the {\em inertial} pump that relies on dynamic asymmetry of the bubble's expansion-collapse cycle when a microheater is located close to one end of a channel~\cite{Yuan1999,Yin2005,Torniainen2012,Kornilovitch2013,Bandopadhyay2014,Zou2015}. Such pumps have characteristic sizes from several microns to hundreds of microns. They can operate even in straight rigid channels of constant cross-section made of robust materials. These highly integrated devices can be made by the batch fabrication processes used in large-scale production of inkjet printheads~\cite{Stasiak2012}.

\begin{figure*}[t]
\begin{center}
\includegraphics[width=0.98\textwidth]{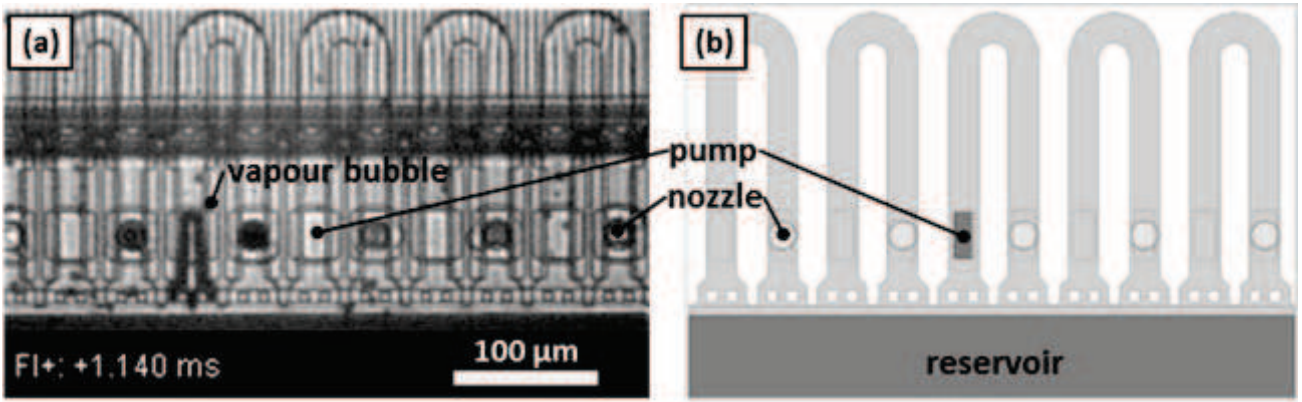}
\end{center}
\caption{(a) A high speed camera image of tested devices with a vapor bubble near maximum excursion. A dark black area in the bottom of the micrograph is a fluidic reservoir, and a grey boundary near it is a ``shelf'' of the channels. Channel width is 22 $\mu$m, channel height (out of the image plane) is 17 $\mu$m. (b) A corresponding plan view. The dark rectangle is a pump microheater. The scale is she same as in panel (a). }   
\label{frdep:fig:tw}
\end{figure*}

We first reported fabrication and characterization of highly integrated inertial pumps in Ref.~\cite{Torniainen2012}. This was followed by a detailed theoretical analysis of the pumping principle in Ref.~\cite{Kornilovitch2013}. The models predicted unusual N-shape flows within 50-$\mu$s long pump pulses. In the first experimental paper \cite{Torniainen2012} this feature was not reported. Resolving it experimentally required high-speed cameras with frame rates of up to a million frames per second and micron-scale spatial resolution. This is the subject of the present work.  

Even without intricate details of bubble generation and interaction in solid-wall tubes~\cite{Zou2015,Ory2000,Kandlikar2006}, the phy\-sics of inertial pumping is complex. Net flow is a result of a subtle imbalance of mechanical momenta of two fluidic columns colliding when the vapor bubble collapses~\cite{Yin2005,Torniainen2012,Kornilovitch2013}. The imbalance is caused by unequal fluid inertia during bubble expansion and asymmetry of vortex formation near channel-reservoir interfaces~\cite{Torniainen2012}. Fluid in the shorter channel arm has lower inertia and reverses its direction earlier. It has more time to accelerate under the external pressure and acquires a larger mechanical momentum by the time of collapse. At collapse, the momentum excess is converted into uniform flow of the entire channel. The latter proceeds in the direction from the short channel arm toward the long one until dissipated by viscous forces. In well-designed pumps, displaced volume per pulse can be as high as 40\% of the maximal bubble volume. At $\sim 10^4$ Hz pulse frequencies, the volumes add up to substantial flow ra\-tes. 

The described mechanism can be easily disturbed by changes to the dynamics of the colliding columns due to dissipation effects, channel morphology and other factors. In order to optimize pump operation, it is important to understand the specifics of expansion, collapse and post-collapse phases of the pump cycle as much as possible. Advancing a previous study~\cite{Torniainen2012},  we report here first successful imaging of {\em single-pulse} inertial flow captured with a 4 $\mu$s resolution. Consistent with theoretical predictions \cite{Torniainen2012,Kornilovitch2013}, we observe a characteristic N-shape flow in the long arm of the channel. The measured flow rate is used to calibrate the full three-dimensional (3D) CFD model \cite{Torniainen2012} and an effective one-dimensional (1D) model \cite{Kornilovitch2013} of inertial pumping to extract the unknown bubble strength. 

Then the {\em frequency dependence} of inertial pumping is studied both experimentally and numerically for the first time. A typical pump cycle lasts $\approx 50$ $\mu$s in our system. As the pump frequency exceeds 20 kHz, flow from a previous pulse is not fully extinguished by the time the next vapor bubble is created. Understanding temporal interaction of vapor bubbles is of fundamental importance. We find that at high frequencies the overall flow rate continues to increase but pump efficiency (displaced volume per pulse) decreases gradually. This behavior has to be taken into account when designing complex microfluidic networks driven by inertial pumps.

\section{\label{frdep:sec:two}
Methods
}

\subsection{\label{frdep:sec:twoone}
Experiment
}

The microfluidic system under study consisted of an array of 528 U-shape channels 22 $\mu$m wide and 17 $\mu$m tall with a linear density of 600 per inch (see figure~\ref{frdep:fig:tw}). Each U-shaped channel had a 33.3 $\times$ 15.0 $\mu$m$^2$ microheater (a thermal inkjet resistor) located near one end of the channel. The channels were fabricated using photoresist (SU8) patterning technique on top of a silicon substrate with pre-fabricated microheaters and CMOS driving electronics~\cite{Stasiak2012}. Both ends of U-channels were fluidically connected to the main fluid reservoir via slots etched through the silicon substrate~\cite{Torniainen2012,Stasiak2012}.

\begin{figure*}[t]
\begin{center}
\includegraphics[width=0.98\textwidth]{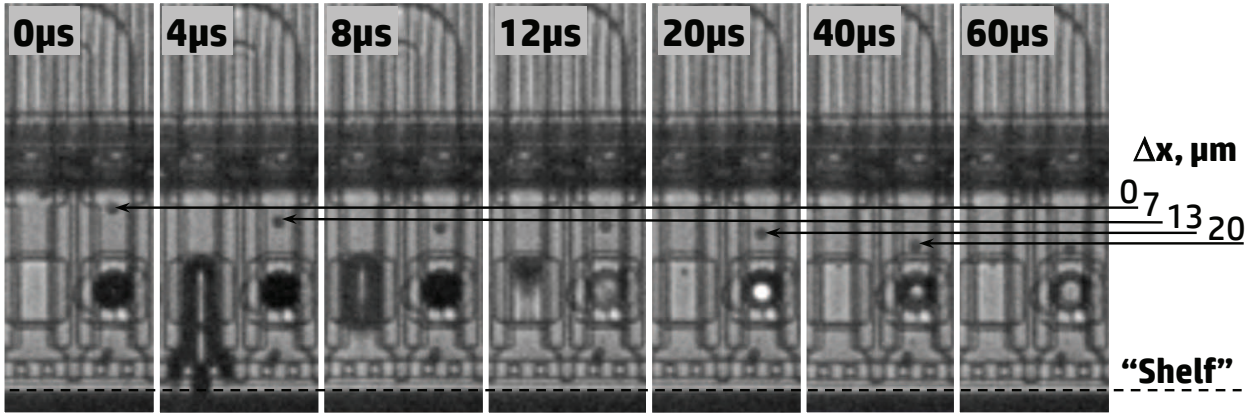}
\end{center}
\caption{An illustrative high-speed image filmstrip showing tracer movement through a channel. Frame times (shown in top left corners) are counted from initiation of the vapor bubble. The vapor bubble is visible in the left leg of the channel in the 4 $\mu$s, 8 $\mu$s and 12 $\mu$s frames.}   
\label{frdep:fig:thtwo}
\end{figure*}

Fabricated devices were filled with a clear test fluid of density $\rho = 1.0$ g/cm$^3$ and viscosity $\mu = 0.86 \pm 0.02$ cP. Viscosity was measured using a Brookfield viscometer at the operating temperature of 45$^{\circ}$C. The total water content was 70-80\%. Microheaters were actuated as described earlier in~\cite{Torniainen2012}. The pump frequency could be varied from a single pulse to 48 kHz. Flow was observed using a conventional optical microscope with coaxial illumination~\cite{Torniainen2012} using a high intensity xenon lamp. Video was recorded with a Vision Research high speed camera Phantom v1610 with frame rates between 250,000 and 1,000,000 frames per second. The exposure (shutter) time of 450 ns was used. The imaging system enabled spatial resolution of $0.8 - 2.5$ $\mu$m, depending on optical magnification used. 

Fluid flows were visualized by 4.0 $\mu$m diameter hollow polystyrene spheres added to the fluid. Particle images were processed frame by frame using the camera time stamp. An example of particle tracing is shown in figure~\ref{frdep:fig:thtwo}. Average particle displacement and net flow were determined based on multiple pulses for every pump frequency tested. Mean displaced volumes and corresponding net flows were calculated by averaging over 20-30 tracers located at different points in the channel cross-section. On average, each channel contained about one tracer particle at any given time. Thus, several dozen pulses had to be processed for every pump frequency. Standard deviation of the particles' displacements across the population was adopted as the displacement error for each frame time. The latter was converted to the displaced volume error and flow rate error shown by error bars in figures~\ref{frdep:fig:thfour} and \ref{frdep:fig:thfive}. 

\begin{figure}[b]
\includegraphics[width=0.48\textwidth]{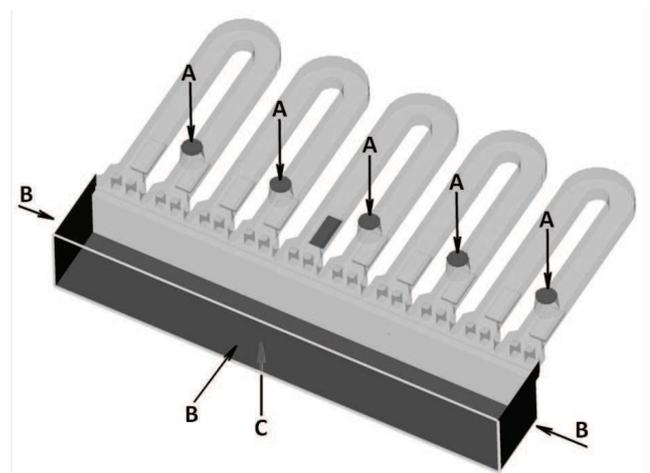}
\caption{Three-dimensional isometric drawing of the device used by the CFD model. Boundary conditions for CFD simulations are set at regions indicated by letters A, B, and C. The black rectangle is the pump.}   
\label{frdep:fig:isoview}
\end{figure}

High-speed photography and particle image velo\-ci\-me\-try (PIV) are powerful tools in microfluidics \cite{Mohammadi2013,Versluis2013}. Prior work on bubble-driven pumps was mo\-st\-ly focused on the bubble evolution as the vapor-fluid interface provides good optical contrast \cite{Yin2005,Geng2001,Wang2004,Sun2009}. Post-collapse flow was also visualized using fluorescent imaging~\cite{Wang2004}. To the best of our knowledge, the present study employs the highest frame rate to-date to measure flow rate over an entire inertial pump cycle.

\subsection{\label{frdep:sec:twotwo}
Computational fluid dynamics (CFD)
}

CFD modeling was used in combination with theory and experiment to gain insight into the behavior of inertial pumps. The first CFD simulations of inertial pumping were reported previously \cite{Torniainen2012}. In this paper, we use the same numerical approach to simulate the physical geometry shown in figure~\ref{frdep:fig:tw}(a). The CFD code, CFD3, is internally developed and is described in detail elsewhere~\cite{Torniainen2012,Tan2015}. CFD3 is very efficient at modeling microscale, inkjet flows using the {\em incompressible} Navier-Stokes equations solved on a staggered, Cartesian grid with piecewise linear interface calculation reconstruction for free surfaces. CFD3 has been used internally for modeling of inkjet droplet ejections for many years and has been extensively validated with experimental data. 

The geometry used in computational analysis is sho\-wn in figure~\ref{frdep:fig:isoview}. Multiple U-shape loops are in the domain and the nozzles (indicated by A in figure~\ref{frdep:fig:isoview}) are open to the atmosphere which is modeled using a gas at atmospheric pressure. The center loop contains a microheater which is used to simulate boiling through a process described in Ref.~\cite{Torniainen2012,Tan2015}. The bottom of the domain (indicated by C in figure~\ref{frdep:fig:isoview}) is modeled using a pressure boundary condition. This pressure boundary condition is atmospheric pressure minus a small back pressure, which is approximately 1000 Pa in this case. All other boundaries have solid walls (indicated by B in figure~\ref{frdep:fig:isoview}). 

The experimental pumping event is simulated by a microheater boiling event and subsequent tracking of the integrated amount of flow through the central U loop. The fluid in the simulation has viscosity $\mu = 0.86$ cP, a density $\rho = 1.0$ g/cm$^3$, a contact angle 45$^\circ$, and surface tension $\sigma = 50$ dyn/cm. The amount of pressure-volume work performed by the boiling event is determined by an internal variable known as critical volume or $V_{cr}$. Critical volume is an adjustable parameter that defines the pressure in the vapor bubble, and subsequently, the amount of work done by the vapor bubble. It is defined in relative units referencing a resistor operating with a standard input pulse on a reference fluid. The value of $V_{cr}$ may vary from 1 if the fluid boils differently than the reference fluid or a non-standard input pulse is used.

\begin{figure}[t]
\includegraphics[width=0.48\textwidth]{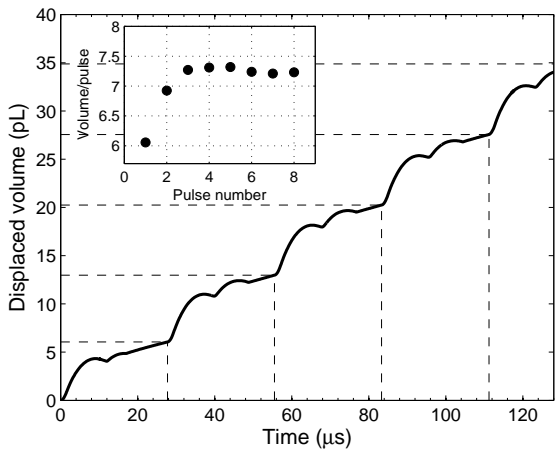}
\caption{A typical CFD run at pump frequency of 36 kHz. Time between pulses is 27.8 $\mu$s. Main panel: the overall time series. Inset: displaced volume as a function of pulse number. The volume per pulse converges to a quasi-stationary limit of 7.2 pL/pulse.}   
\label{frdep:fig:CFDtimeseries}
\end{figure}

In contrast to the previous work~\cite{Torniainen2012}, here we simulate multiple successive pump cycles over long continuous runs. Each subsequent bubble is injected into remnant flow from the previous cycle. After several pulses the system reaches a quasi-stationary state with displaced volume being independent of pulse number. This is a more accurate way of assessing interaction between pulses. In \cite{Torniainen2012}, pump frequency effects were estimated by truncating single pulse flows at appropriate times, the single pulse always starting from a state of rest.    

A typical CFD3 pumping result for multiple pulses is shown in figure~\ref{frdep:fig:CFDtimeseries}. The cumulative amount of fluid volume passing through the central U-loop is measured and shown as a function of time in the main panel. The next step is to calculate displaced volume per pulse as shown in the inset graph. Typically, within a few pulses, the displaced volume approaches a quasi-stationary va\-lue, which is 7.2 pL/pulse in this example.

\begin{figure}[t]
\includegraphics[width=0.48\textwidth]{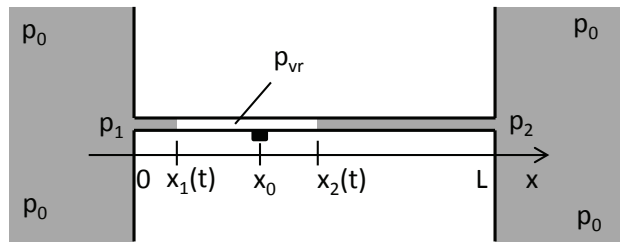}
\caption{Simplified 1D schematic of an inertial pump. The small black rectangle marks the location of a microheater $x_0$. $x_0$ is also the initial positions for {\em both} vapor-fluid interfaces at the start of bubble expansion. $p_{0}$ is the external pressure equal in both reservoirs. The residual vapor pressure $p_{\rm vr} = 0.3$ atm is assumed to be independent of $x_0$. $p_1$ and $p_2$ are the dynamic pressures at the channel-reservoir interfaces and are inflow-outflow asymmetric.}   
\label{onedpt2:fig:one}
\end{figure}

\subsection{\label{frdep:sec:twothree}
The one-dimensional model
}

A simplified one-dimensional model of inertial pumping was developed previously~\cite{Kornilovitch2013}. The basic geometry is shown in figure~\ref{onedpt2:fig:one}. A pump state is completely described by positions of two vapor-fluid interfaces $x_1(t)$ and $x_2(t)$ and their temporal derivatives $\dot{x}_1(t)$ and $\dot{x}_2(t)$. In terms of dimensionless time $\tau$ and interface coordinates $\xi_{1,2}$, cf. table~\ref{frdep:tab:one}, the dynamic equations for the left and right fluidic columns read
\begin{eqnarray}
      \xi_1   \, \xi''_1 + \frac{1}{2} \xi'^2_1 H( \xi'_1) + 
             \beta \, \xi_1   \, \xi'_1       & = &          1 \: ,
\label{onedf:eq:twelve} \\
( 1 - \xi_2 ) \, \xi''_2 - \frac{1}{2} \xi'^2_2 H(-\xi'_2) + 
             \beta \, ( 1 - \xi_2 ) \, \xi'_2 & = &        - 1 \: .
\label{onedf:eq:thirteen}
\end{eqnarray}
Here the prime denotes a derivative with respect to $\tau$, $\beta$ is effective friction, and $H(z)$ is a Heaviside step function: $H(z > 0) = 1$, and $H(z < 0) = 0$. The pressures at the channel-reservoir interfaces are given by $p_{1,2} = p_0 - \frac{1}{2} \rho \, \dot{x}^2_{1,2} H( \pm \dot{x}_{1,2})$. The interface pressures are equal to the external pressure $p_0$ during expansion, when the fluid is ejected into reservoirs. During collapse, the pressure is reduced by a Bernoulli term~\cite{Yuan1999}. Using CFD3, we have verified that the flow-asymmetric dynamic pressure is more representative of the inertial pump physics than the constant boundary conditions also considered in Ref.~\cite{Kornilovitch2013}. After the two columns collide at point $\xi_c$ at time $\tau_c$, the entire channel fluid moves as a whole with a constant mass. Fluid motion in this {\em post-collapse} phase is described by an equation
\begin{equation}
\xi'' + \frac{1}{2} \, \xi'^2 \, {\rm sgn}(\xi') + \beta \, \xi' = 0 \: ,
\label{onedf:eq:twentyone}
\end{equation}
with $\xi(\tau_c) = \xi_c$ and $\xi'(\tau_c) = \eta_c$. The differential equations (\ref{onedf:eq:twelve})-(\ref{onedf:eq:twentyone}) can be solved by direct integration in time domain using standard numerical methods.

\begin{table}
\renewcommand{\tabcolsep}{0.2cm}
\renewcommand{\arraystretch}{1.5}
\begin{center}
\begin{tabular}{|c|c|c|}
\hline\hline
 Variable/parameter   &  Formula                              & Values                  \\  \hline
\hline 
 Time                 &  $\tau = \frac{t}{L} \sqrt{\frac{p_0-p_{\rm vr}}{\rho}}$  
          &  $0 - 5$                                                                    \\  \hline
 Interface position   &  $\xi_{i} = \frac{x_{i}}{L}$          & $( 0.0 - 1.0 )$         \\  \hline
 Interface velocity   &  $\eta_{i} = \frac{\partial \xi_{i}}{\partial \tau} = \xi'_{i}$ 
          &  $0 - 1$                                                                    \\  \hline
 Microheater location &  $\xi_0 =  \frac{x_{0}}{L}$           & $( 0.0 - 1.0 )$         \\  \hline         
 Bubble strength      &  $\alpha = \frac{q_0}{\rho A L} \sqrt{\frac{\rho}{p_{0} - p_{\rm vr}}}$
          &  $0.1 - 1.0$                                                                \\  \hline 
 Effective friction   &   
       $\beta = \frac{\kappa \mu L}{\rho A} \sqrt{\frac{\rho}{p_0-p_{\rm vr}}}$       
          &  $1 - 10$                                                                   \\  \hline  
\hline 
\end{tabular}
\end{center}
\caption{
Dimensionless variables and parameters used in the effective 1D model of inertial pumping~\cite{Kornilovitch2013}. Here $t$ is physical time and $x_{1,2}$ are the physical locations of vapor-liquid interfaces. $x_0$ is the physical location of the microheater center, $L$ is the channel length, $A$ is the channel cross-sectional area, $\rho$ and $\mu$ are the fluid density and viscosity, $p_0 = 10^5$ Pa is the atmospheric pressure, and $p_{\rm vr} = 0.3 \cdot 10^{5}$ Pa is the residual pressure of the vapor bubble. Quantity $q_0$ is the initial liquid momentum that is approximately given by the product of vapor pressure, channel cross-section and duration of the high-pressure phase. $\kappa$ is the geometrical loss factor that depends on the shape of channel cross-section. $\kappa = 29.34$ for a $22/17$ rectangle (the present case) \cite{Timoshenko1970}. The last column lists values typical for the experimental systems studied in this work.  
} 
\label{frdep:tab:one}
\end{table}

The initial conditions for the expansion-collapse ph\-a\-se are 
\begin{equation}
\xi_1(0) = \xi_2(0) = \xi_0 \: , 
\label{onedtwo:eq:two}
\end{equation}
\begin{equation}
\xi'_1(0) = - \frac{\alpha}{\xi_0}     + \eta_0 \: , \hspace{0.5cm} 
\xi'_2(0) =   \frac{\alpha}{1 - \xi_0} + \eta_0 \: .
\label{onedf:eq:seventeen}
\end{equation}
where $\alpha$ is the dimensionless bubble strength, $\xi_0$ is the microheater location, and $\eta_0$ is the velocity prior to bubble generation. Dimensionless model parameters are defined in table~\ref{frdep:tab:one}. Note that the starting point of bubble expansion $\xi_0$ (which coincides with the microheater location) can be {\em any} value on the interval $(0,1)$, only excluding end points 0 and 1. Mathematically, the dynamic model is well defined even in the limit $\xi_0 \rightarrow 0$. A small fluid mass results in a large deceleration that immediately turns around the flow, so that the singular point $\xi = 0$ is never reached. Clearly, in real devices $\xi_0$ is limited by finite sizes of the microheater ($20 - 50$ $\mu$m in length) and of the microchannel ($L = 200 - 2000$ $\mu$m in length). Thus, in practical calculations the starting point $\xi_0 = x_0/L$ typically lies within the interval $0.01 < \xi_0 < 0.99$.     

A non-zero starting velocity $\eta_0$ is a {\em novel} feature of the 1D theory compared with the single-pulse case studied in Ref.~\cite{Kornilovitch2013}. If a new pulse is initiated before flow from the previous pulse is completely extinguished, the new vapor bubble will be initiated in a moving fluid. This changes the expansion-collapse dynamics of both fluidic columns and may affect overall net flow. The pre-bubble velocity $\eta_0$ has a one-to-one correspondence with the pulse frequency $f$. In practical calculations, $\eta_0$ is set to some value, then expansion-collapse flow is solved to obtain post-collapse velocity $\eta_c$, and finally post-collapse flow is solved to find the moment when flow slows down from $\eta_c$ to $\eta_0$. The procedure determines a total time after which the system returns to its original state. An inverse of this time, expressed in physical units, is the pump frequency that can be directly compared with experiment.       

We close this section by comparing the benefits of numerical and theoretical methods. CFD provides the most accurate treatment of an inertial flow problem, and is typically the first method to try on a new system. It takes into account flow features in reservoirs, finite microheater dimensions, realistic vapor pressures and other elements of a real architecture. However, simulations take a long time: runs reported in this work take up to several days to complete. In contrast, the 1D model equations can be solved in a few seconds, but they do not capture the system elements listed above. The 1D model always requires calibration with both experiment and CFD to make sure the model parameters are realistic and in correspondence with practical devices. Once the key parameters are set (for example the unknown bubble strength $\alpha$), the 1D model becomes an excellent tool for exploring the vast space of microfluidic architectures. We believe that at this stage of inertial pump development, it is necessary to carry both approaches as they are complementary.

\begin{figure}[t]
\begin{center}
\includegraphics[width=0.48\textwidth]{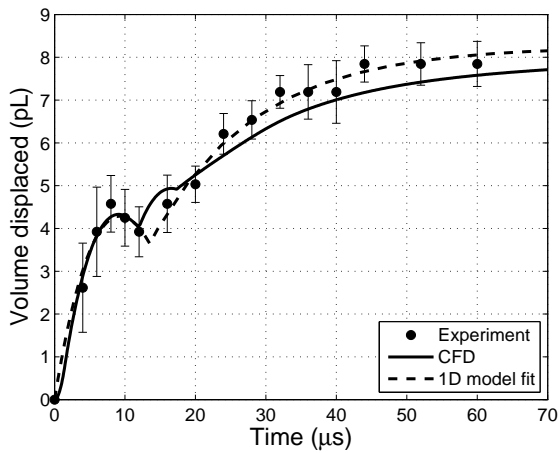}
\end{center}
\vspace{0.2cm}
\caption{Single-pulse flow curve. Circles with error bars are experimental data. The solid line is CFD flow for $V_{cr} = 1.1$. The dashed line is the best fit by the 1D model with parameters $\xi_0 = 0.158$, $\alpha = 0.276$. The experimental error bars represent standard deviations of the displacement of about 30 tracer particles located at random points of the channel cross-section.}   
\label{frdep:fig:thfour}
\end{figure}

\section{\label{frdep:sec:three}
Results
}

First, we characterized individual pumping events at low pump frequencies below $10$ Hz. At such frequencies, successive pulses do not overlap in time. Volume displaced through the long channel arm versus time is plotted in figure~\ref{frdep:fig:thfour}. Experimental data points are shown by circles. Note that the curve has a characteristic N-shape. Once the vapor bubble is created, the fluid is ``kicked'' through the channel by high vapor pressure. Once the vapor cools off and the pressure drops (within a couple of $\mu$s), the flow starts to decelerate under the combined action of friction forces and a negative pressure difference. At around 8 $\mu$s, the long arm flow reverses direction and the bubble collapses at about 12 $\mu$s. During collapse the long arm has a smaller mechanical momentum than the short arm (for the reasons explained in Refs.~\cite{Torniainen2012,Kornilovitch2013}) and the flow reverses direction again, now from negative to positive. This secondary, post-collapse flow lasts for another 40-50 $\mu$s until it is fully extinguished by friction forces and the entire fluid comes to rest. Note that the N-shape flow was predicted theoretically~\cite{Torniainen2012,Kornilovitch2013}, but it has been observed experimentally for the first time here.    

Results from the CFD simulations are presented in figure~\ref{frdep:fig:thfour}. Comparisons to experimental data were used to determine the critical volume $V_{cr} = 1.1$. The fluid parameters used in the simulations were described in section~\ref{frdep:sec:twotwo}. The CFD simulations match the experiment quite well during the initial phase until the bubble collapses at about 12 $\mu$s. Then there is a small deviation between the experiment and numerical model. The small hump in displaced volume in the CFD simulation from 14 to 18 $\mu$s is due to bubble rebound. This rebound is seen in our experiment (and it was observed in the past \cite{Zhao2000,Glod2002}) but its effect on displaced volume is masked by the volume measurement error.

\begin{figure}[t]
\includegraphics[width=0.48\textwidth]{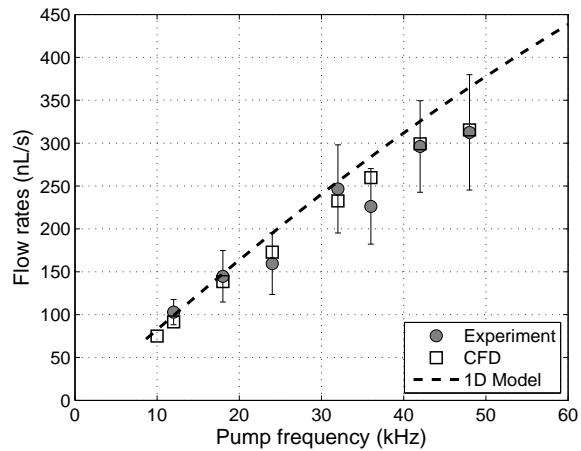}
\caption{Experimental flow rates at different frequencies superimposed on predictions of CFD (for $V_{cr} = 1.1$) and the 1D model with parameters extracted from single-pulse fits.}   
\label{frdep:fig:thfive}
\end{figure}

The best 1D model fit of single-pulse flow is shown in figure~\ref{frdep:fig:thfour} by a dashed line. The overall agreement is good, except that the total displaced volume comes out slightly larger than the measured value (8.2 pL vs 7.8 pL) but still within experimental error bars. In fitting, the channel length (403 $\mu$m), width (22 $\mu$m), height (17 $\mu$m) as well as fluid density $\rho = 1.0$ g/cm$^3$ and viscosity $\mu = 0.86$ cP were set to experimental values. The microheater location $\xi_0 = x_0/L$ and effective bubble strength $\alpha$ were left as adjustable parameters. The optimal values have been found to be $\xi_0 = 0.158$ and $\alpha = 0.276$. The $\alpha$ value falls right within the estimate given in table \ref{frdep:tab:one}. $\xi_0 = 0.158$ is about 10\% larger than the ``theoretical value'' of 0.140. The latter is obtained if the $x_0$ is associated with the geometrical center of the pump microheater, which is 56.5 $\mu$m away from the reservoir, cf. figure~\ref{frdep:fig:tw}. The remaining differences can be attributed to a finite size of the microheater, presence of open nozzles and particle filters that are not captured within the 1D model.  

The flow rate dependence on pump frequency is now discussed. Quasi-steady flow rates were measured at $f = 12$, 18, 24, 32, 36, 42, and 48 kHz. Experimental data points are shown in figure~\ref{frdep:fig:thfive} by circles. Since the duration of one pulse is about 50 $\mu$s, successive pulses do not overlap in time until $f$ exceeds 20 kHz. Thus, the flow rate is expected to be linear in $f$ for $f < 20$ kHz. At higher frequency, the flows begin to interfere with each other. It is of fundamental interest to know the sign of the effect. Experimental data suggest that the interaction is negative, i.e. the flow rate grows with the pump frequency sub-linearly.

\begin{figure}[t]
\includegraphics[width=0.48\textwidth]{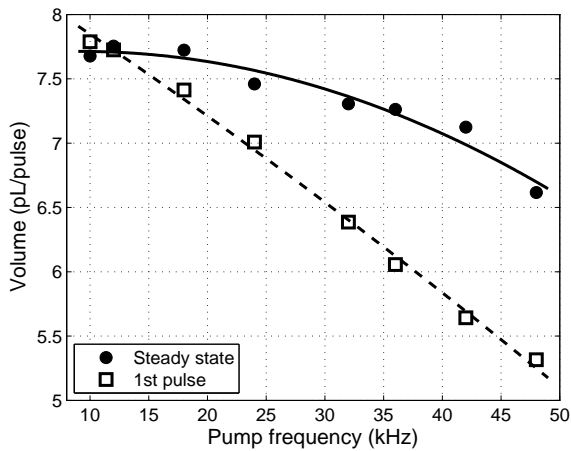}
\caption{Steady-state volume per pulse compared with first-pulse volume for different pump frequencies. The lines are parabolic fits to guide the eye.}   
\label{frdep:fig:firstpulse}
\end{figure}

The sublinear behavior at high frequencies is reproduced by both CFD and 1D models as shown in figure~\ref{frdep:fig:thfive}. In both cases, model parameters were taken from the single pulse fits described above. In the CFD case, the run time was long enough to reach the quasi-stationary flow, as explained earlier. There is excellent agreement between CFD and experiment, within experimental error bars. The 1D model slightly overestimates the flow at high frequencies. We attribute this to be a consequence of the single-pulse fit that produced a displaced volume slightly higher than experiment, cf. figure~\ref{frdep:fig:thfour}. That may have led to a slightly larger effective bubble strength and less interaction between successive pulses as a result. Overall agreement between our models and experiment is good.      

Physically, the reduction of flow rate is caused by the new vapor bubble disrupting post-collapse flow generated by the previous pulse. For a well-designed pump, post-collapse flow provides the bigger portion of the overall displaced volume, and the new bubble cuts short the contribution of this secondary phase to the previous pulse. As a result, the volume {\em per pulse} gradually goes down. It is instructive to compare the effect of such short-cutting on both first-pulse and steady state flows, which is done in figure~\ref{frdep:fig:firstpulse}. The first pulse occurs in a stagnant flow and the corresponding displacement is defined by the moment when the single-pulse flow curve of figure~\ref{frdep:fig:thfour} is terminated by the second pulse. However, subsequent pulses take place in nonzero flow, which increases the overall displacement for the same time interval between pulses. As a result, some volume lost at first pulse is recovered as the flow approached steady state. That is why the steady state curve shown in figure~\ref{frdep:fig:firstpulse} is more flat that the first-pulse curve.

\section{\label{frdep:sec:four}
Summary
}

Due to its small size and complete integrability, the inertial pump may become an essential element of future fluidic integrated circuits. It is important to understand various aspects of this new technology. In this paper, single-pulse dynamics has been imaged by a high-speed camera for the first time. This enabled calibration of the full three-dimensional CFD model and an effective one-dimensional model with respect to the unknown bubble strength. Then, variation of time-average flow rate with pump frequency was investigated, also for the first time. Both experiment and models have shown sublinear behavior at high frequencies. Such behavior is caused by temporal interaction of successive pump pulses, when each new pulse disrupts the post-collapse flow generated by the previous pulse. As a result, the overall flow rate still grows at high frequencies but the displaced volume per pulse goes down. From this qualitative picture, the sublinear dependence of the total flow rate should be more prominent in systems with longer post-collapse flows, e.g. in microchannels with larger cross-sectional areas than those studied in this work.

\begin{acknowledgements}

The authors wish to thank T. Deskins and M. Brown for high speed videos, K. Vandehey, M. Monroe, M. Regan, C. Macleod, T. Mattoon and P. Stevenson for general support of this work. 

\end{acknowledgements}



\end{document}